\documentclass[amsmath,amssymb,twoside,twocolumn,british,aps,pra,superscriptaddress]{revtex4}
\usepackage[T1]{fontenc}
\usepackage[latin1]{inputenc}
\usepackage{graphicx}

\makeatletter

\usepackage{babel}
\makeatother
\begin{document}
\newcommand {\ve}[1]{\mbox{\boldmath $#1$}}
\newcommand \beq{\begin{equation}}
\newcommand \beqa{\begin{eqnarray}}
\newcommand \beqann{\begin{eqnarray*}}
\newcommand \eeq{\end{equation}}
\newcommand \eeqa{\end{eqnarray}}
\newcommand \eeqann{\end{eqnarray*}}
\newcommand \ga{\raisebox{-.5ex}{$\stackrel{>}{\sim}$}}
\newcommand \la{\raisebox{-.5ex}{$\stackrel{<}{\sim}$}}

\title{\textmd{Low-temperature crossover in the momentum distribution of
cold atomic gases in one dimension}}

\author{Vadim V. Cheianov}

\affiliation{NORDITA, Blegdamsvej 17, DK-2100 Copenhagen, Denmark}

\author{H. Smith}

\author{M. B. Zvonarev}

\affiliation{\O rsted Laboratory, H. C. \O rsted Institute,
Universitetsparken 5, DK-2100 Copenhagen, Denmark}

\begin{abstract}
The momentum distribution function for the two-component 1D gases
of bosons and fermions is studied in the limit of strong
interatomic repulsion. A pronounced reconstruction of the
distribution is found at a temperature much smaller than the Fermi
temperature. This new temperature scale, which equals the Fermi
temperature divided by the dimensionless coupling strength, is a
feature of the two-component model and does not exist in the
one-component case. We estimate the parameters relevant for the
experimental observation of the crossover effect.
\end{abstract}

\maketitle

\section{Introduction}

During the last few years the great advances in the art of cooling
and trapping atomic gases have made it possible to
study experimentally the behavior of fermion and boson gases in
lower dimensions\cite{Gorlitz}. Of particular interest are the
quasi-one-dimensional systems which can be realized as
one-dimensional (1D) tubes in a two-dimensional optical lattice
\cite{Tolra,Richard,P-04,Kohl}. The ground states of such systems are characterized
by the ratio between the interaction energy and the kinetic
energy. At high densities, the kinetic energy dominates and the
system can be described by mean-field theory. At low densities,
one enters a regime of strong coupling where the interaction
energy dominates the kinetic energy. The difference between this
strongly interacting regime and the regime of weak interactions is
more than just quantitative. One should expect that strong
interactions result in phenomena absent in the weakly interacting
case. The question of what these phenomena are and their
experimental consequences is essential for our understanding of
low-dimensional systems and, more generally, the physics of
strongly interacting matter.

It is evident that the presence of internal degrees of freedom enriches
the phase diagram of a physical system and may even give rise to
completely new phenomena.
One-dimensional systems in general and one-dimensional quantum gases in particular,
are no exceptions to this rule. In this paper we discuss a
phenomenon in quantum gases which is specific to the strongly interacting
regime and the presence of internal degrees of freedom.
We show that, due to the emergence of a soft propagating mode
in the limit of strong repulsion, one-dimensional
gases (both bosonic and fermionic) undergo a dramatic
reconstruction of the one-particle momentum distribution as the
temperature is varied on a very small scale. This
temperature scale is proportional to the propagation velocity of
the soft mode and vanishes in the limit of infinitely strong
interparticle repulsion.

The paper is organized as follows. In Section \ref{1C2C} we give
a qualitative comparison of one- and two-component quantum gases
in 1D and introduce the relevant energy scales. We
show that in the two-component case there exists a new low-temperature
regime where the behavior of the quantum gas should be
qualitatively different from that at zero temperature.
Sections \ref{OPDM1} and \ref{OPDM2} are technical. In Section
\ref{OPDM1} we present an exactly solvable model, which grasps
the physics discussed in Section \ref{1C2C}. We give an exact
representation for the one-particle density matrix in this model.
In Section \ref{OPDM2} we analyze short- and long-distance asymptotics
of the density matrix and discuss the numerical recipe  for evaluating it
at intermediate distances. Section \ref{LTR} contains the main result
of the paper: The effect of the low temperature reconstruction of
the momentum distribution function is demonstrated for both
boson and fermion gases. In Section \ref{Discussion} we discuss the possibility of
observing this reconstruction
effect experimentally and estimate the magnitude of the coupling
strength for such an experiment in terms of the parameters
characterizing highly elongated harmonic traps.

\section{One-component versus two-component gases in one dimension
\label{1C2C}}

 One-dimensional {\it uniform} gases with {\it
zero-range} interaction serve as a test ground for a wide class of
one-dimensional systems. This was realized long ago by Lieb and
Liniger, who used the Bethe-ansatz method \cite{KBI}
to obtain the exact wave
functions and spectrum of the one-component boson gas~\cite{LL-63}.
They considered the following Hamiltonian for $N$ identical bosons
of mass $m$ and chemical potential $\mu$,
\begin{equation}
H=-\frac{\hbar^2}{2m}\sum_{i=1}^N\frac{\partial^2}{\partial
x_i^2}+g_{\rm 1D}\sum_{1\le i<j\le N}\delta(x_i-x_j)-\mu N.
\label{LL2}
\end{equation}
The interaction constant $g_{\rm 1D}$ has the dimension of energy
times length and is conventionally written in terms of a
characteristic length $a_{\rm 1D}$ according to
\begin{equation}
g_{\rm 1D}=\frac{2\hbar^2} {ma_{\rm 1D}}.
\label{gdef}
\end{equation}
The dimensionless coupling strength
$\gamma$ involves the number of particles per unit length, $n_{\rm
1D}=N/L$, where $L$ is the length of the 1D system, and is given
by \beq \gamma=\frac{mg_{\rm 1D}}{\hbar^2n_{\rm
1D}}=\frac{2}{n_{\rm 1 D}a_{\rm 1D}}.\label{gamma} \eeq Thus when
the density decreases, the coupling strength increases, and one
approaches the limit of impenetrable bosons, the so-called
Tonks-Girardeau gas \cite{TG-36}. In this limit the thermodynamic
properties of the system are the same as those of a
one-dimensional gas of non-interacting one-component fermions with
the Fermi energy \beq E_{\mathrm F}=\pi^2\frac{\hbar^2n_{\rm
1D}^2}{2m}.\label{my}\eeq Henceforth we set $\hbar=2m=1.$ These
constants will be restored in Section~\ref{Discussion}.

Similarly to the one-component case, one can consider a {\it
two-component} gas of fermions (bosons). Such gases consist of
particles with some internal quantum parameter $\sigma$, which can
take two values. For brevity we refer to this parameter as
``spin'', and label the corresponding states by $\sigma=\uparrow,
\downarrow.$ In the situation where the interaction constant
$g_{\rm 1D}$ is tuned to be independent of the ``spin'' index, we
take the first-quantized Hamiltonian of the system to be the same
as for the one-component case, Eq.~\eqref{LL2}. The information
about the system being two-component is encoded in the symmetry of
the coordinate wave function. To introduce the ``spin'' indices in
the Hamiltonian explicitly one can recast it in the second
quantized form using the particle creation and annihilation
operators $\psi^{\dagger}_\sigma(x)$ and $\psi_\sigma (x).$
Written in the second-quantized form the interaction term in
Eq.~\eqref{LL2} is
\begin{equation}
\frac{g_{\rm 1D}}2 \int dx\, n_{\uparrow}(x)
n_{\downarrow}(x) \label{fermiint}
\end{equation}
for two-component fermions, and
\begin{equation}
\frac{g_{\rm 1D}}4 \int dx\, :n(x)^2: \label{Boseint}
\end{equation}
for two-component bosons.  Here
$n_\sigma(x)=\psi_\sigma^\dagger(x) \psi_\sigma(x)$ is the density
of particles of type $\sigma;$ the total density $n(x)$ is
$n(x)=n_\uparrow(x)+n_\downarrow(x),$ and the symbol $::$ stands
for the normal ordering. For interactions of this form the
total number of particles $N$ and the total number
$N_{\uparrow(\downarrow)}$ of ``spin'' up (down) particles are
good quantum numbers. Note, that the operator
\eqref{fermiint} is the most generic form of contact
interaction for fermions, which is a consequence of the Pauli
principle. In the bosonic case, however, the operator \eqref{Boseint}
represents a special situation where the three coupling constants
are tuned to be equal.

The density $n_{1\mathrm D}$ is controlled by the chemical
potential $\mu,$ Eq.~\eqref{LL2}. In the infinite $\gamma$ limit
the relation between $n_{1\mathrm D}$ and $\mu$ does not depend on
the statistics of particles and has the following simple form
\begin{equation}
{\pi n_{1\mathrm D}}={\sqrt \mu}. \label{density}
\end{equation}
For the one-component gas this result can be found in {\it e.g.}
Refs.~\cite{TG-36}, for the two-component gas in
Ref.~\cite{Izergin}. It follows from Eq.~\eqref{density} that
$\sqrt \mu$ has a dimension of inverse length. Hereafter we
measure distances in units of $1/\sqrt\mu.$

What temperature scales characterize 1D gases with the Hamiltonian
\eqref{LL2} in the limit of strong repulsion, $\gamma\gg 1$? For
the one-component system there is only one temperature scale: the
Fermi temperature $T_{\rm F}=E_{\rm F}/k_{\rm B}$, where $k_{\rm
B}$ is the Boltzmann constant. Low temperatures $T\ll T_{\rm F}$
correspond to the degenerate quantum gas. The inclusion of
``spin'' gives rise to another characteristic temperature $T_0\ll
T_F,$ defined by the bandwidth for the ``spin'' excitations. This
bandwidth can be estimated as follows. Consider the sector of the
Hilbert space with a given number of particles $N$ and a
polarization $M=N_\uparrow-N_\downarrow.$ Denote by $E(M)$ the
ground state energy of the Hamiltonian \eqref{LL2} in this sector.
For fermions the function $E(M)$ has its absolute minimum at $M=0$
\cite{Yang-67} in agreement with the Lieb-Mattis theorem
\cite{LM-62}. For bosons the function $E(M)$ reaches its absolute
minimum at $|M|=N$ \cite{Li-03}, giving rise to the demixing
transition \cite{CaHo-2003}. The energy difference per particle
between the polarized (demixed) and unpolarized (mixed) phases
\begin{equation}
\epsilon=\frac{\vert E(N)-E(0)\vert}{N} \label{epsilondef}
\end{equation}
gives an estimate for the bandwidth of ``spin'' excitations: $T_0
\sim \epsilon/k_{\mathrm B}.$ In the large $\gamma$ limit $\epsilon$ is
easily found from the thermodynamic Bethe ansatz \cite{Yang-67}.
For fermions one finds that to the leading order in ${1}/{\gamma}$
\begin{equation}
\epsilon=\frac{8\ln 2}3  \frac{E_{\mathrm F}}{\gamma}.
\label{gripper}
\end{equation}
One can see that the characteristic temperature scale for ``spin''
excitations is given by $T_0=T_{\rm F}/\gamma.$ The same estimate
is valid for bosons.

The two scales $T_0$ and $T_{\rm F}$ are well separated when
$\gamma$ is large enough. In this situation one can distinguish
between two different quantum (which means $T\ll T_{\rm F}$)
regimes: $T< T_0$ and $T> T_0.$ The first regime has been studied
extensively and is conventionally described by the Luttinger
Liquid (LL) theory \cite{GNT-98}; the results on the second one
appeared very recently \cite{CZ-04I,CZ-04II,BF-04} for the
limiting case $T_0\ll T\ll T_{\rm F}$ and are qualitatively
different from the LL theory predictions. This can be understood
by noticing that for $T>T_0$  ``spin'' degrees of freedom are
strongly ``disordered'' thus violating the LL theory applicability
conditions. At the same time, the density degrees of freedom are
not affected by the temperature until it becomes of the order of
$T_{\rm F}.$

\section{One-particle density matrix: an exact representation
\label{OPDM1}}

In this Section we investigate the one-particle density matrix of
the two-component strongly repulsive ($\gamma\to+\infty$)
1D gases with the Hamiltonian \eqref{LL2} for the two limiting
cases $T\ll T_0$ and $T_0\ll T$ of the two quantum regimes
discussed in Section~\ref{1C2C}. Our considerations will be based
on the exact solution of the model Hamiltonian \eqref{LL2} in the
infinite $\gamma$ limit. The two different regimes $T\ll T_0$ and
$T_0\ll T$ are accessed by choosing the appropriate order of
limits $T\to 0$ and $\gamma \to \infty,$ as described in
\cite{CZ-04I}: To ensure $T_0\ll T$ one should take the limit
$\gamma\to\infty$ first and then set $T=0$ in the wave functions
and spectra of the model, while for the LL regime $T\ll T_0$ one
should take the limit $T\to 0$ first.

For two-component systems the one-particle density matrix is
defined as
\begin{equation}
\rho(x)=
\frac12\left[\langle\psi^\dagger_\uparrow(x)\psi_\uparrow(0)\rangle
+
\langle\psi^\dagger_\downarrow(x)\psi_\downarrow(0)\rangle\right].
\label{corrdef2}
\end{equation}
Since the interaction terms Eqs.~\eqref{fermiint} and
\eqref{Boseint} are invariant with respect to ``spin'' rotation,
one has $\langle\psi^\dagger_\uparrow(x)\psi_\uparrow(0)\rangle =
\langle\psi^\dagger_\downarrow(x)\psi_\downarrow(0)\rangle.$ The
density of particles is $n_{1\mathrm D}=2\rho(0).$ Both physical
intuition and crucial technical advance come from the observation
that in the infinite $\gamma$ limit the exact Bethe-ansatz
many-particle eigenfunctions of the Hamiltonian \eqref{LL2}
factorize into a coordinate part which is similar to the wave
function of non-interacting spinless fermions and a spin part
which corresponds to an eigenstate of the isotropic Heisenberg
chain. Using this factorization Ogata and Shiba \cite{Ogata}
derived an explicit formula for the density matrix
\eqref{corrdef2} in the case of two-component fermions on a
lattice (the Hubbard model). The two-component fermion gas being
the limit of the Hubbard Hamiltonian for the vanishing filling
factor, Ogata and Shiba's formula can be used. It also has a
straightforward extension to the bosonic case. Both bosonic and
fermionic cases are discussed in the paragraph below.

Using the results of Ref.~\cite{Ogata} we write the density matrix
$\rho(x)$ as
\begin{equation}
\rho(x)=\frac12\langle \chi^{\dagger}(x) \omega(\mathcal N_x)
\chi(0) \rangle_{\mathcal F}, \label{OS}
\end{equation}
where the average $\langle\rangle_{\mathcal F}$ is taken over the
ground state of the system of non-interacting one-component
fermions with creation and annihilation operators $\chi^{\dagger}$
and $\chi,$ respectively. The density of these fermions coincides
with the density of physical particles $n_{1\mathrm D}.$ The
operator
\begin{equation}
\mathcal N_x=\int_0^x dx' \chi^{\dagger}(x') \chi(x')
 \label{Ndef}
\end{equation}
counts the number of particles between points $0$ and $x$.
Assuming lattice regularization with an infinitesimal lattice
constant, $\mathcal N_{x}$ is a Hermitian operator with integer
eigenvalues. To define the function $\omega(N)$ consider  an
infinite isotropic Heisenberg chain. Its Hilbert space  is spanned
by vectors $\vert \dots\sigma_1\sigma_2\dots\sigma_j\dots\rangle$,
where $\sigma_j=\uparrow, \downarrow$ is the spin at the $j$-th
site. In this space one may introduce a string operator
$\Omega_N,$ which acts as a cyclic shift operator on the string of
spins of length $N$:
\begin{eqnarray}
\Omega_N \vert \ldots\sigma_j\sigma_{j+1} \ldots
\sigma_{j+N-2}\sigma_{j+N-1}\sigma_{j+N}\dots\rangle =\nonumber\\
\vert \ldots\sigma_{j+N-1}\sigma_{j}\sigma_{j+1}\ldots
\sigma_{j+N-2}\sigma_{j+N}\ldots\rangle.
\end{eqnarray}
The function $\omega(N)$ is defined as follows
\begin{equation}
\omega(N)=e^{i (\pi-\phi) N} \langle \Omega_N
\rangle_{\mathcal{H}}. \label{omegadef}
\end{equation}
Here $\phi$ is a statistical angle: $\phi=0$ for bosons and
$\phi=\pi$ for fermions. In the case $T\ll T_{0}$ the average
$\langle\rangle_{\mathcal H}$ in Eq.~\eqref{omegadef} is taken
over the ground state of the isotropic Heisenberg chain, which is
antiferromagnetic for the fermion gas and ferromagnetic for the
boson gas. In the case $T\gg T_{0},$ the average
$\langle\,\rangle_{\mathcal H}$ is to be taken over all possible
spin configurations.

For the boson gas at exactly zero temperature and both fermion and
boson gases at $T\gg T_{0}$ the calculation of $\omega(N)$ is
relatively simple. For bosons at zero temperature the average
$\langle\,\rangle_{\mathcal H}$ in Eq.~\eqref{omegadef} is taken
over the fully polarized ferromagnetic ground state of the spin
chain which gives
\begin{equation}
\omega(N)=e^{i \pi N}. \label{Omegabosons}
\end{equation}
For bosons and fermions at $T\gg T_{0}$ the average
$\langle\,\rangle_{\mathcal H}$ in Eq. \eqref{omegadef}  runs over
all spin configurations. This gives
\begin{equation}
\omega(N)=e^{i (\pi-\phi) N} 2^{-N} \label{omegaht}.
\end{equation}
In both equations Eqs.~\eqref{Omegabosons} and \eqref{omegaht} the
function $\omega(N)$ is an exponential of $N:$ $\omega(N) \sim e^{
\lambda N},$ for which the average \eqref{OS} can be expressed in
terms of the Fredholm determinant. More precisely, one obtains the
following representation for the density matrix \eqref{corrdef2}:
\begin{equation}
\frac{\rho(x)}{\rho(0)}=\det(\hat I+\hat V_1+\hat
V_2)(x)-\det(\hat I+\hat V_1)(x). \label{detrep}
\end{equation}
Here the determinant
\begin{eqnarray}
\det (\hat I +\hat V) = \sum_{N=0}^{\infty}\frac1{N!}
\int_{-1}^{1}{\rm d} k_1\ldots\int_{-1}^{1}{\rm d} k_N \nonumber\\
\times\det\left[
\begin{matrix}
V(k_1,k_1)& \cdots & V(k_1,k_N)\cr \vdots & \ddots & \vdots \cr
V(k_N,k_1)& \cdots & V(k_N,k_N)
\end{matrix}
\right] \label{det}
\end{eqnarray}
is the Fredholm determinant of a linear integral operator $\hat V$
with the kernel defined on
$[-1,1]\times[-1, 1].$ In our
case the kernels $V_1$ and $V_2$ are:
\begin{eqnarray}
V_1(k,p)=\alpha\frac{\sin\frac{x}2(k-p)}{\pi(k-p)}\label{V1def}\\
V_2(k,p)=\frac12 \exp\{-i\frac{x}2(k+p) \} \label{V2def}
\end{eqnarray}
The values of the parameter $\alpha$ are given in the Table~\ref{alphatable}.
\begin{table}
\begin{center}
\begin{tabular}{|r|c|}
\hline
&\hskip.5cm$\alpha$\hskip.5cm \vphantom{c}\\
\hline
\hskip.5cm one-component boson gas,\hfill $T=0$ &$-2$\\
\hskip.5cm two-component boson gas,\hfill $T\ll T_0$ &$-2$\\
\hskip.5cm two-component boson gas, \hfill $T\gg T_{0}$ &$-3/2$\\
\hskip.5cm two-component fermion gas, \hfill $T\ll T_{0}$ &---\\
\hskip.5cm two-component fermion gas, \hfill $T\gg T_{0}$ &$-1/2$\\
\hline
\end{tabular}
\end{center}
\caption{The values of the parameter $\alpha$ entering
Eq.~\eqref{V1def} are given.
 For the two-component fermion
gas at $T\ll T_{0}$ no exact results on $\rho(x)$ are known.}
\label{alphatable}
\end{table}
Recall that we measure distances in units of $1/\sqrt \mu$, Eq.~\eqref{density}

It was mentioned above Eq.~\eqref{Omegabosons} that the ground
state of the two-component bosons is fully polarized in ``spin''
\cite{Li-03}. Therefore, at $T\ll T_0$  the bosons
become effectively one-component (compare first and second lines
in the Table~\ref{alphatable}) and we turn back to the original
Lieb-Liniger model~\cite{LL-63}. The determinant representation
\eqref{detrep} in this model was obtained by Lenard~\cite{L-64}.
For the two-component fermion gas at $T\gg T_{0}$
Eq.~\eqref{detrep} was obtained in Ref.~\cite{BL-87}. Izergin and
Pronko calculated both bosonic and fermionic density matrices in
Ref.~\cite{Izergin}.

\section{One-particle density matrix: asymptotics and numerics \label{OPDM2}}

In this Section we analyze the short and long-distance asymptotics
of the determinant representations of the density matrix
\eqref{corrdef2}. We also describe the numerical procedure which
deals with the intermediate length scales. The results of this
Section will be used for the calculation of the momentum
distribution function given in the next Section.

Since $\rho(x)=\rho(-x),$ we assume $x\ge0.$ For {\it short
distances} ($x\ll 1$) one can easily find from Eqs.~\eqref{detrep}
and~\eqref{det} that
\begin{equation}
\frac{\rho(x)}{\rho(0)}=
1-\frac{x^2}6-\frac{\alpha}\pi\frac{x^3}{18}+
\frac{x^4}{120}+\frac{\alpha}\pi\frac{11x^5}{2700}+\ldots .
\label{rhosmall}
\end{equation}
For $\alpha=0$ this is simply the expansion of $\sin x/x$. Notice that
higher order terms that are not exhibited on the r.h.s.
of Eq.~\eqref{rhosmall}, depend
on $\alpha$ nonlinearly.
Next, consider the {\it long-distance} expansion of Eq.~\eqref{detrep}.
Compared to the short-distance analysis, this is a more sophisticated task.
For the boson gas at $T\ll T_0$ the analysis carried out
in Refs.~\cite{VT-79} shows the power-law decay of the density matrix:
\begin{equation}
\frac{\rho(x)}{\rho(0)}=\frac{C}{\sqrt
x}\left[1+\frac1{8x^2}\left(\cos{2x}-\frac14\right)+\frac{3}{16x^3}\sin2x\right]
\end{equation}
with  relative corrections of the order of $x^{-1}.$ The constant
$C$ is given by  $C=\pi e^{1/2} 2^{-1/3} A^{-6}$ with $A=1.2824271\ldots$ being
Glaisher's constant. For the two-component fermion gas at $T\gg T_0$ the decay
is exponential~\cite{BL-87}:
\begin{equation}
\rho(x)=C_1 e^{-2\nu  x} x^{\Delta_{\mathrm F}} \sin(2 x-\nu\ln x
+C_2). \label{rhofermi}
\end{equation}
Here $C_1$ and $C_2$ are some constants which are given explicitly
in Ref.~\cite{B-91}, while
\begin{equation}
\Delta_{\mathrm F}=\frac{\nu^2}2-1, \quad \nu=\frac{\ln2}\pi. \label{fermiexp}
\end{equation}
The relative corrections to Eq.~\eqref{rhofermi} are of the order
of $x^{-1}.$ The technique developed in the
papers~\cite{CZ-04I,CZ-04II} and \cite{BF-04} for the
two-component fermion gas can be easily adopted to the case of the
two-component boson gas at $T\gg T_0$. Like for fermions, the density
matrix decays exponentially in the large $x$ limit
\begin{equation}
\rho(x)= \mathrm{const}\; e^{-2\nu x}x^{\Delta_\mathrm B}.
\label{long-exp}
\end{equation}
The anomalous exponent
\begin{equation}
\Delta_\mathrm B=\frac{1}{2}(\nu^2-1)
\end{equation}
and $\nu$ is given by Eq.~\eqref{fermiexp}.
The relative correction to Eq.~\eqref{long-exp} is of the order of
$x^{-1}.$

Finally, we turn to intermediate values of $x.$ To analyze the
representation~\eqref{detrep}, we use instead of Eq.~\eqref{det}
an alternative definition of the Fredholm determinant
\cite{Smirnov}. Let $V$ be an $L\times L$ matrix with the entries
$V_{ab}=V(k_a,k_b).$ Let
\begin{equation}
k_a=\left(\frac{2a}{L-1}-1\right), \quad a=0,1,\ldots,L-1.
\end{equation}
Then the Fredholm determinant Eq.~\eqref{det} can be represented as follows:
\begin{equation}
\det(\hat I+\hat V)=\lim\limits_{L\to\infty}
\det\nolimits_L\left(I+\frac{2}{L-1} V\right).
\label{detfinite}
\end{equation}
The matrix $I$ on the right hand side of the equation
\eqref{detfinite} is the $L\times L$ identity matrix.
We use the representation
\eqref{detfinite} to calculate $\rho(x)$ for $x$ in the
range $0\le x\le 20.$ For the figures shown in the next Section
we used the number of divisions $L=800.$

\section{Low-temperature reconstruction of the momentum distribution
\label{LTR}}
In this Section we present our results on the
momentum distribution function of quantum two-component gases in
two temperature regimes $T\ll T_0$ and $T\gg T_0$ discussed in
Section~\ref{1C2C}.

The momentum distribution function $n(k)$ for a one-dimensional
gas is given by
\begin{equation}
n(k)=\int dx e^{-ikx}\rho(x), \label{Gasympt}
\end{equation}
where $\rho(x)$ is the one-particle density matrix \eqref{corrdef2}.
Recall that $x$ is measured in units of
$1/\sqrt \mu$ as defined below Eq.~\eqref{density}.

\begin{figure}
\begin{center}
\includegraphics[clip,width=8cm]{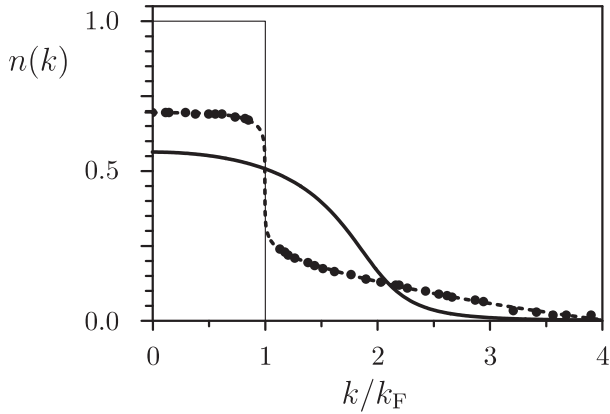}
\end{center}
\caption{The momentum distribution function $n(k)$ for
two-component fermions. The thick solid curve is our result for $T
\gg T_0.$ The dashed curve corresponds to $T\ll T_0$ and is
obtained by the interpolation of the data (filled circles) from Ref.~\cite{Ogata}.
The thin solid line shows the
distribution function for non-interacting two-component fermions. The momentum $k$
is normalized to the Fermi momentum $k_{\mathrm F}=\pi n_{1\mathrm D}/2.$}
\label{fig_2}
\end{figure}

First, consider the gas of two-component fermions.
The momentum distribution for different
temperature regimes is shown in Fig.~\ref{fig_2}.
The $T\ll T_0$ curve was obtained by the interpolation of
the data given in Ref.~\cite{Ogata} for the Hubbard model at
low filling. The $T\gg T_0$ curve was
calculated using the determinant representation discussed in Sections
\ref{OPDM1} and \ref{OPDM2}.
One can see a dramatic change in the shape of the momentum
distribution function as the temperature increases from $T\ll T_0$ to $T\gg T_0.$ The
singularity of the momentum distribution at $k=k_F=\pi n_{1\mathrm D}/2,$ predicted
by the Luttinger theorem \cite{Luttinger}, disappears and the momentum distribution
spreads out to larger $k.$ Another very pronounced effect is that
the large momentum tail of the momentum distribution gets strongly
suppressed as the temperature increases. This
is a counterintuitive result from the point of view of physics of weakly
interacting systems, where the fraction of high energy particles
grows monotonously with increasing temperature. In strongly
correlated systems, however, bare particles have no overlap
with elementary excitations of the system and the momentum
distribution of bare particles is not directly related to the
distribution of energy between the eigenmodes of the system.

\begin{figure}
\begin{center}
\includegraphics[clip,width=8cm]{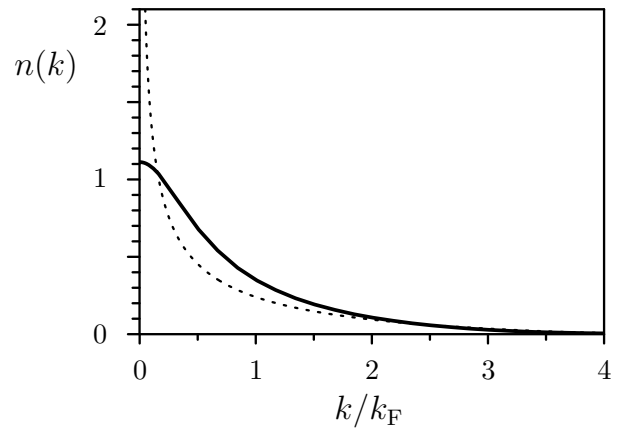}
\end{center}
\caption{The momentum distribution function $n(k)$ for two-component
bosons. The solid curve is the result for $T\gg T_0,$ the dashed curve
is for $T\ll T_0.$ For convenience we normalize the momentum $k$ to the
Fermi momentum $k_{\mathrm F}=\pi n_{1\mathrm D}/2$ of a fermion system
with the same density.} \label{fig_1}
\end{figure}

Next, consider the gas of two-component bosons. Using the results
of Sections \ref{OPDM1} and \ref{OPDM2} we calculate the momentum distribution
function Eq.~\eqref{Gasympt} numerically for both $T\ll T_0$
and $T\gg T_0.$
The results of the calculations are shown in
Fig.~\ref{fig_1}.
One can see that the momentum distribution in the ground state of
bosons at $T\ll T_0$ is divergent
at $k=0.$ This behavior takes place at any
coupling strength $\gamma$ \cite{Popov-80,Haldane-81}
and is the manifestation of the quasi-long-range order in the system,
that is, quasicondensate.
The situation is different at $T\gg T_0.$ Due to the exponentially
decaying term in the asymptotic expression Eq.~\eqref{long-exp},
the function $n(k)$ is continuous with all its derivatives for all
$k$. The suppression of the quasicondensate component
as the temperature increases from $T\ll T_0 $ to above $T_0$
has a simple physical explanation: due to the excitation of the soft
degrees of freedom the system loses coherence and cannot exhibit
quasi-long-range order. For the large momentum tail of $n(k)$
one gets from Eq. \eqref{rhosmall}:
\begin{equation}
n(k) \sim -\frac{\alpha}{k^4}, \qquad k\to \infty.
\end{equation}
Comparing the values of $\alpha$ given in Table \ref{alphatable} one
can see that the fraction of particles with large momentum
decreases as the temperature increases from $T\ll T_0$ to above $T_0$
similarly to the fermionic case.

\section{Discussion \label{Discussion}}

One-dimensional quantum gases were recently created and
studied in an optical lattice forming an array of well separated
one-dimensional tubes \cite{Tolra}. The particle momentum
distribution in such systems can be observed in experiments
involving ballistic expansion \cite{Richard,P-04,Kohl}.
Large values of the coupling constant, $\gamma\sim 200,$
were reported in Ref.~\cite{P-04}. Such values of $\gamma$ would
give a two orders of magnitude separation between the two
temperature scales $T_0$ and $T_{\mathrm F},$ which is sufficient
for the momentum reconstruction to be observable. There are, however,
certain geometric limitations on the validity of the model, Eq.~\eqref{LL2},
considered in this paper, which we discuss below. For the convenience of
the reader we restore the constants $\hbar$ and $m$ in the equations.

Consider an elongated harmonic trap of
transverse frequency $\omega_{\perp}$ and axial
frequency $\omega_z=\lambda\omega_{\perp},$ $\lambda\ll1,$ containing
a quantum gas of $N$ particles.
The characteristic length $a_{1\mathrm D}$ in Eq.~\eqref{gdef}
is related to the three-dimensional scattering length $a$ and the
transverse width $a_{\perp}=\sqrt{\hbar/m\omega_{\perp}}$ by \cite{Ol}
\begin{equation}a_{\rm 1D}=
\frac{a_{\perp}^2}{a}\left(1-C_1\frac{a}{a_{\perp}}
\right),
\label{rm1D}
\end{equation} where
$C_1=-\zeta(1/2)/\sqrt{2}=1.0326\ldots$ is a numerical constant,
and $\zeta(x)$ is the Riemann zeta function.
Since the local chemical potential $\mu(z)$  is given by
$\mu(z)=\pi^2\hbar^2n_{\rm 1D}^2(z)/2m$, the one-dimensional
density is found within the Thomas-Fermi approximation  to be
$n_{\rm 1D}(z)=(2N-z^2/a_z^2)^{1/2}/{\pi a_z}$, where
$a_z=\sqrt{\hbar/m\omega_z}$. The value of $\gamma,$ Eq.~\eqref{gamma},
may then be expressed in terms of the particle number $N$,
the scattering length $a$ and the trap
frequencies $\omega_{\perp}$ and $\omega_z$. At the trap center
($z=0$) one obtains the simple expression
\begin{equation}
\label{gamma0} \gamma(0) =\pi\left(\frac{2}{\lambda N}
\right)^{\frac 1 2} \frac{a}{a_{\perp}-C_1 a}.
\end{equation}
From Eq.~\eqref{gamma0} it is clear that to achieve large values
of $\gamma(0)$ at fixed $N$ and $\lambda$
the system must be tuned to the resonance $a_{\perp}-C_1 a=0.$
However, very close to
the resonance the scattering amplitude will no longer correspond
to a delta-function potential in real space due to the presence of
higher-order terms in the relative momentum. This may be
demonstrated from the expansion of the scattering amplitude $f$ as
given in \cite{Ol}, \beq f=-\frac{1}{1+ika_{\rm
1D}-i\sqrt{2}\zeta(3/2)(ka_{\perp})^3/8}. \eeq From this
expression one notes that the terms of third order in $k$ are
negligible only when $a_{\rm 1D}/a_{\perp}\gg
\sqrt{2}\zeta(3/2)(ka_{\perp})^2/8$. If we identify $k$ with the
Fermi wave number $k_{\rm F}$ which has its maximum value $k_{\rm
F}=\pi n_{1D}(0)=\sqrt{2N}/a_z$ in the center ($z=0$) we obtain
the condition \beq a_{\rm 1D}\gg
a_{\perp}\frac{\sqrt{2}\zeta(3/2)}{4}N\lambda \eeq which in combination
with Eq.~\eqref{rm1D} implies that $(a_{\perp}/a-C_1)/N\lambda \gg 1,$
in order for our model to be applicable. Combining this
result with Eq.~\eqref{gamma0} we arrive at the upper bound
for the interaction constant $\gamma$ for the trap with given
$\lambda$ and $N$:
\begin{equation}
1\ll\gamma(0)\ll \pi\sqrt 2 \left(\frac{1}{\lambda N} \right)^{\frac 3 2}.
\end{equation}
For traps with $\lambda \sim 300$ \cite{Kohl}
containing $N=30$ particles this estimate gives $\gamma\ll 140.$

\section{Conclusions}
In conclusion, the momentum distribution of two component boson and
fermion gases was considered in the limit of strong interatomic
repulsion. It was shown that due to the strongly correlated nature
of the system the presence of the internal degrees of freedom of
the atoms results in the emergence of a new low temperature
decoherent state absent in one-component gases. The onset of this
state with increasing temperature is marked by a pronounced
change of the momentum distribution of the atoms.

\section*{Acknowledgement}
The authors would like to thank J. Dalibard for helpful discussions.
M.B.~Zvonarev's work was supported by the Danish Technical
Research Council via the Framework Programme on Superconductivity.

\end{document}